\documentclass[english,aps,manuscript,endfloats,superscriptaddress]{revtex4}
\usepackage{graphicx}
\usepackage{epsfig}
\usepackage{amsmath}
\usepackage{graphics}
\begin{document}                              

\author{Maxim F. Gelin}
\affiliation{Department of Chemistry,           
         Technische Universit\"at M\"unchen
         D-85747 Garching, Germany}
\author{Michael Thoss}
\affiliation{Institute of Theoretical Physics and Interdisciplinary Center for Molecular Materials,           
         Friedrich-Alexander-Universit\"at Erlangen-N\"urnberg, 
Staudtstr.\ 7/B2,
         D-91058 Erlangen, Germany}

\title{Thermodynamics of a subensemble of a canonical ensemble}
\begin{abstract}
Two approaches to describe the thermodynamics of a subsystem that
interacts with a thermal bath are considered. Within the first approach, 
the mean system energy $E_{S}$ is identified with the expectation
value of the system Hamiltonian, which is evaluated with respect to
the overall (system+bath) equilibrium distribution. 
Within the second approach, 
the system partition function $Z_{S}$ is considered as the fundamental 
quantity, which is postulated to be the
ratio of the overall (system+bath) and the bath partition functions,
and the standard thermodynamic relation $E_{S}=-d(\ln Z_{S})/d\beta$
is used to obtain the mean system energy.
Employing both classical and quantum mechanical treatments, 
the advantages and shortcomings of
the two approaches are analyzed in detail for various different systems.
It is shown that already within classical mechanics both approaches 
predict significantly different results for thermodynamic quantities
provided  the system-bath interaction is not bilinear
or the system of interest consists of more than a single particle.
Based on the results, it is concluded that the first approach is superior.

\end{abstract}
\maketitle

\section{Introduction}

Let us consider a system in equilibrium at  fixed volume and temperature
described by a canonical ensemble.
If  the partition function $Z$ for the corresponding
canonical distribution is known, all thermodynamic quantities, such as
the internal energy, the entropy, and the specific heat can be calculated
 by simple
differentiations of $\ln Z$ with respect to the temperature $T$.
In many situations it is of interest to consider a subsystem of the 
overall system, e.g.
a smaller system that interacts with its environment (in the following referred to as the bath). The question arises, how to describe
the thermodynamics of the subsystem?


If the system-bath coupling is weak, then (to leading (zero) order
in the system-bath coupling) the system can be described by  a canonical
distribution determined by the corresponding system Hamiltonian. 
This finding builds the basis of standard statistical thermodynamics. If the
coupling is not weak, however, the interaction with the 
bath affects the system density matrix,
the system partition function $Z_{S}$, and all  thermodynamic
quantities \cite{weiss,opp89,gev00,nie02,han05}. 
Furthermore, new interesting effects arise 
due to the system-bath entanglement \cite{nie02,han05,jor05,for06,mah07,nie00,lutz09}. 
In such a case, it is not straightforward, e.g., to define the energy of the system unambiguously. Different definitions
are possible which incorporate, to a certain extent, the system-bath
interaction into the system energy. 

There exist two major approaches to describe the thermodynamics
of a subsystem (strongly) coupled to a bath. The first approach 
(in the following referred to as approach I) considers the mean energy of
the system as fundamental quantity and assumes that it is given by
the expectation value of the system Hamiltonian which is evaluated
with respect to the total (system+bath) canonical equilibrium distribution
\cite{roux,han06,han08}. The second approach (approach II), on the other hand, 
is based on the partition function of the system, $Z_{S}$, and postulates 
it to be given as the ratio of the total (system+bath)
and the bath partition functions. Based on the thus defined partition function of the system, the standard relations of
thermodynamics are invoked to calculate the internal
energy, the entropy, and the specific heat by differentiations
of $\ln Z_{S}$ with respect to the temperature $T$ \cite{for85,for88,for07}. 

In recent work \cite{nie02,han06,han08}, it was shown that the two
approaches give different results for the specific heat of
a quantum mechanical
point particle or a harmonic oscillator bilinearly coupled to a harmonic
bath, despite the fact that the results are identical in the classical case.
In the present work, we analyze the two approaches and assess their 
validity in detail for several
different systems including several quantum and classical point particles and 
nonlinear
system-bath coupling. 


Our main findings are summarized as follows:
If approach I is employed, then the knowledge of the system partition function
$Z_{S}$ alone is not sufficient to describe the thermodynamics of the system
and the standard procedure calculating the thermodynamic quantities
by differentiations of $\ln Z_{S}$ with respect to $T$ is no valid. 
Additional knowledge of a bath-induced interaction
operator $\Delta_{S}$ (see below) is required.  We derive general 
expressions for the
mean system energy, entropy, and the specific heat in terms of $Z_{S}$
and $\Delta_{S}$. 

If, on the other hand, the strategy of approach II is followed to define
the partition function of the system $Z_S$, 
then obtaining the internal energy,
the entropy, and the specific heat by differentiations of $\ln Z_{S}$
with respect to $T$ cannot be justified by referring to the standard
thermodynamic machinery. The so-obtained expressions must be regarded
as the definitions of the corresponding quantities, whose validity
must be proven \textit{a posteriori}. 

We also show that the differences between the two approaches are not of
purely quantum origin. The approaches predict significantly different results
already within classical
mechanics, provided the system-bath interaction is not bilinear
and/or the system of interest consists of more than a single particle.
The general results are illustrated by explicit calculations of
thermodynamic quantities for several classical and quantum model systems.
Based on the thus obtained results, we conclude  approach I is 
clearly superior to approach II, both on physical and logical grounds.

\section{Thermodynamics of a canonical ensemble}

Let us consider a canonical ensemble which is kept in  constant
volume $L$ at  fixed temperature $T.$ The corresponding equilibrium distribution
(density matrix) reads 
\begin{equation}
\rho=Z^{-1}\exp\{-\beta H\},\,\,\, Z=\textrm{Tr}(\exp\{-\beta H\}).\label{ro}\end{equation}
Here $H$ is the Hamiltonian, $Z$ is the partition function, $\beta\equiv 1/(k_{B}T)$
($k_{B}$ being the Boltzmann constant), and $\textrm{Tr}(...)$ denotes
the integration over  phase space variables in  case of a classical ensemble or
taking the trace in  case of a quantum ensemble, respectively. 
We further define the (information)
ensemble entropy operator as \begin{equation}
S\equiv-k_{B}\ln\rho=\frac{1}{T}(H-F)\label{S}\end{equation}
and the free energy \begin{equation}
F\equiv-\frac{1}{\beta}\ln Z,\label{F}\end{equation}
so that the canonical distribution (\ref{ro}) can alternatively be
written as \begin{equation}
\rho=\exp\{-\beta(H-F)\}.\label{ro1}\end{equation}
Averaging Eq.\ (\ref{S}) over the canonical distribution (\ref{ro1})
we obtain the following expression for the ensemble-averaged entropy
\begin{equation}
\left\langle S\right\rangle =\frac{1}{T}(\left\langle H\right\rangle -F),\label{S1}\end{equation}
where we have used the notation 
$\left\langle ...\right\rangle \equiv\textrm{Tr}(\rho...)$.
The specific heat is determined as \begin{equation}
C=\frac{d\left\langle H\right\rangle }{dT}.\label{C}\end{equation}
Using Eqs. (\ref{ro}) and (\ref{ro1}) we can express the internal
energy, the entropy, and the specific heat through the free energy:\begin{equation}
\left\langle H\right\rangle =-\frac{d(\ln Z)}{d\beta}=F-T\frac{dF}{dT},\label{Hd}\end{equation}
\begin{equation}
\left\langle S\right\rangle =-\frac{dF}{dT},\label{Sd}\end{equation}
\begin{equation}
C=T\frac{d\left\langle S\right\rangle }{dT}=-T\frac{d^{2}F}{dT^{2}}.\label{Cd}\end{equation}
We emphasize that Eqs.\ (\ref{Hd})-(\ref{Cd}) are a direct consequence
of the explicit form and temperature-dependence of the canonical operator
(\ref{ro}). 

The second law of thermodynamics in differential form then reads
\begin{equation}
Td\left\langle S\right\rangle =d\left\langle H\right\rangle .\label{2}\end{equation}

\section{Thermodynamics of a reduced canonical ensemble}

\subsection{General expressions }\label{genexpr}

The  expressions for the thermodynamical observables of a canonical ensemble
listed above can be found in any textbook on statistical
thermodynamics. To study the thermodynamics of a subensemble let us now 
consider a system that is interacting with
its environment (in the following referred to as the bath). 
We assume that the Hamiltonian $H$ can be
split into the system ($S$) Hamiltonian, the bath ($B$) Hamiltonian,
and their coupling, 
\begin{equation}
H=H_{S}+H_{B}+H_{SB}.\label{H1}\end{equation}
Here $H_{S}$ depends solely on the system degrees of freedom and $H_{B}$
depends exclusively on the bath degrees of freedom. 

To calculate an observable which depends on the
system degrees of freedom only, it is expedient to introduce the reduced
 density matrix of the system, $\rho_{S}$, which is defined
by averaging the total density matrix $\rho$ over the bath degrees
of freedom
\begin{equation}
\rho_{S}\equiv Z^{-1}\textrm{Tr}_{B}\left(\exp\{-\beta(H_{S}+H_{B}+H_{SB})\}\right).\label{rS}\end{equation}
Following Kirkwood \cite{kir}, Eq.\ (\ref{rS}) can equivalently be
rewritten as \cite{roux} 
\begin{equation}
\rho_{S}=Z_{S}^{-1}\exp\{-\beta(H_{S}+\Delta_{S})\},\,\,\, Z_{S}\equiv Z.\label{rSa}
\end{equation}
Here, we have defined  the bath-induced interaction
operator 
\begin{equation}
\Delta_{S}\equiv-\frac{1}{\beta}\ln\textrm{Tr}_{B}\left(\exp\{-\beta(H_{S}+H_{B}+H_{SB})\}\right)-H_{S},\label{Del}
\end{equation}
which, in general, depends on the system degrees of freedom and on
the temperature, $T$. 

If the system-bath coupling is week ($\left\Vert H_{SB}\right\Vert \ll\left\Vert H_{S}\right\Vert ,\left\Vert H_{B}\right\Vert $)
then, to the leading (zero) order in the system-bath coupling, the interaction operator is given by $\Delta_{S}=-\ln (Z_{B})/\beta$.
Here $Z_{B}$ is the partition function of the bath canonical distribution\begin{equation}
\rho_{B}=Z_{B}^{-1}\exp\{-\beta H_{B}\},\,\,\, Z_{B}\equiv\textrm{Tr}_{B}\left(\exp\{-\beta H_{B}\}\right)\label{Zb}\end{equation}
In this weak-coupling case, the overall partition function factorizes, 
$Z=Z_{S,c}Z_{B}$ (the subscript ``c'' stands for ``canonical'') 
and the reduced density matrix (\ref{rS})
is determined by the canonical distribution for the system alone
\begin{equation}
\rho_{S,c}=Z_{S,c}^{-1}\exp(-\beta H_{S}),\,\,\, Z_{S,c}=\textrm{Tr}_{S}(\exp\{-\beta H_{S}\}).\label{rS1}\end{equation}
In general, however, $H_{SB}$ cannot be neglected in Eq.\ (\ref{Del}) and
the simple expression (\ref{rS1}) is not valid. 

Analogously to Eq. (\ref{S}), we introduce the system free energy
\begin{equation}
F_{S}\equiv-\frac{1}{\beta}\ln Z_{S},\label{FS}\end{equation}
so that the distribution (\ref{rSa}) can equivalently be rewritten
as 

\begin{equation}
\rho_{S}=\exp\{-\beta(H_{S}+\Delta_{S}-F_{S})\}.\label{rSaF}\end{equation}

The distributions (\ref{rSa}) and (\ref{rSaF}) can be used for calculating
the expectation value of any operator $Y_{S}$ which depends only on the
system degrees of freedom \cite{foot3}, 
\begin{equation}
\left\langle Y_{S}\right\rangle \equiv\left\langle 
Y_{S}\right\rangle _{S},\,\,\,\left\langle ...\right\rangle _{S}
\equiv\textrm{Tr}_{S}(\rho_{S}...)\equiv\textrm{Tr}_{S+B}(\rho...).
\label{Av}
\end{equation}

\subsection{Approach I (Mean energy approach)}

In  approach I,  the mean system  
energy $E_{S}$ is associated with the expectation value of the 
system Hamiltonian
$\left\langle H_{S}\right\rangle $ \cite{roux,han06,han08}. Accordingly,
the system contribution to the specific heat is determined as
\begin{equation}
C_{S}=\frac{d\left\langle H_{S}\right\rangle }{dT}.\label{CS}\end{equation}

To obtain an expression for the entropy of the system, we 
follow the presentation in Sec.\ 2 and define the information system
entropy as 
\begin{equation}
S_{S}\equiv-k_{B}\ln\rho_{S}=\frac{1}{T}(H_{S}+\Delta_{S}-F_{S}).\label{SS}
\end{equation}
Averaging the expression over distribution (\ref{rSa}), we obtain the
analogue of Eq. (\ref{S1})
\begin{equation}
\left\langle S_{S}\right\rangle =\frac{1}{T}(\left\langle H_{S}\right\rangle 
+\left\langle \Delta_{S}\right\rangle -F_{S}).\label{S1S}
\end{equation}
In passing, we note that there exists a controversy in the literature 
whether the so 
defined $\left\langle S_{S}\right\rangle$ can indeed be considered as
a proper thermodynamic entropy of the reduced system \cite{nie02,oco07,but05}. 
For the purpose of this paper, 
Eqs.\ (\ref{SS}) and  (\ref{S1S}) can merely be considered as intermediate 
mathematical expressions, which allow us to conveniently derive
 the formulas discussed below.

In contrast to the situation discussed in Sec.\ 2, the system free energy
$F_{S}$  is no longer a universal object 
which determines all 
relevant thermodynamic quantities $\left\langle S_{S}\right\rangle $,
$\left\langle H_{S}\right\rangle $, and $C_{S}$ through Eqs. (\ref{Hd})-(\ref{Cd}).
It is the explicit dependence of $\Delta_{S}$ on the system degrees
of freedom and the temperature which violates the standard thermodynamic
expressions for the reduced system. It is straightforward, however, 
to generalize Eqs.\ (\ref{Hd})-(\ref{Cd}) if we assume that operators $H_{S}$
and $\Delta_{S}$ commute. In this case, using Eqs.\
(\ref{FS})-(\ref{CS}), we obtain the expressions
\begin{equation}
-\frac{d(\ln Z_{S})}{d\beta}=F_{S}-T\frac{dF_{S}}{dT}=\left\langle H_{S}\right\rangle +\left\langle \Delta_{S}\right\rangle -T\left\langle \frac{d\Delta_{S}}{dT}\right\rangle ,\label{alla}
\end{equation}
\begin{equation}
\left\langle S_{S}\right\rangle =\left\langle \frac{d\Delta_{S}}{dT}\right\rangle -\frac{dF_{S}}{dT},\label{allb}
\end{equation}
\begin{equation}
C_{S}=T\frac{d\left\langle S_{S}\right\rangle }{dT}+\left\langle \frac{d\Delta_{S}}{dT}\right\rangle -\frac{d\left\langle \Delta_{S}\right\rangle }{dT}.\label{allc}
\end{equation}
It is important to note that 
\begin{equation}
\left\langle \frac{d\Delta_{S}}{dT}\right\rangle \neq\frac{d\left\langle \Delta_{S}\right\rangle }{dT}\label{Ne}
\end{equation}
due to the explicit temperature-dependence of $\rho_{S}$. If $H_{SB}\neq0$,
then the last two terms in Eqs. (\ref{alla}) and (\ref{allc}) give
a non-negligible contribution. If the bath-induced potential $\Delta_{S}$
in Eq. (\ref{allb}) is temperature-independent, then the usual formula
(\ref{Sd}) for the entropy holds, but the expressions for
$\left\langle H_{S}\right\rangle $
(\ref{alla}) and $C_{S}$ (\ref{allc}) do contain the bath-induced
contributions. Furthermore, Eq. (\ref{allc}) shows that
the second law of thermodynamics (in differential form) is modified
to \begin{equation}
Td\left\langle S_{S}\right\rangle =d\left\langle H_{S}\right\rangle -\left\langle d\Delta_{S}\right\rangle +d\left\langle \Delta_{S}\right\rangle .\label{2a}\end{equation}

To summarize, knowing the partition function $Z_{S}$ or the free
energy $F_{S}$ alone is not enough to calculate 
$\left\langle S_{S}\right\rangle $,
$\left\langle H_{S}\right\rangle $, and $C_{S}$ within  approach I. 
Instead, the more general Eqs.\ (\ref{alla})-(\ref{2a}) must be used.
 This should be taken into account if, e.g.,
work theorems are employed to obtain the system partition functions 
beyond the weak system-bath coupling limit \cite{jar04,gel08}. 
The general expressions (\ref{alla})-(\ref{2a}) are also important 
for the thermodynamics of  small systems  \cite{FQMT,rub07,rit07}.

Eqs.\ (\ref{alla})-(\ref{2a}) have been derived assuming 
that $H_{S}$ and $\Delta_{S}$ commute. 
This requirement is  not as restrictive as it might seem 
at first glance and is enough for the purposes of the present article. 
It is obviously fulfilled within classical mechanics. 
It is also fulfilled in the semiclassical limit considered below 
described within the Wigner function formalism, 
provided we start from the total (system+bath) Wigner distribution 
and introduce 
the reduced Wigner distribution of the system by averaging the total 
Wigner distribution 
over the bath degrees of freedom (see Sec.\ IV C).
   
Even in the general case, where $[H_{S},\Delta_{S}]\neq0$, analogues
of Eqs. (\ref{Hd})-(\ref{Cd}) can be derived. 
To this end, we first rewrite the system density matrix
in the form
\begin{equation}
\rho_{S}=Z_{S}^{-1}\exp\{-\beta H_{S}\}\exp\{-\beta\widetilde{\Delta}_{S}\}\label{rSa1}
\end{equation}
with a slightly redefined bath-induced interaction
operator 
\begin{equation}
\widetilde{\Delta}_{S}\equiv-\frac{1}{\beta}\ln \left( \exp\{\beta H_{S}\}\textrm{Tr}_{B}\left(\exp\{-\beta(H_{S}+H_{B}+H_{SB})\}\right)\right).\label{Dela}
\end{equation}
Second, we replace $H_{S}\rightarrow\gamma H_{S}$ in Eq. (\ref{rSa1}) 
(with $\gamma$ being a numerical parameter)
but keep $\widetilde{\Delta}_{S}$ unchanged, 
so that the distribution (\ref{rSa1}) becomes 
\begin{equation}
\rho_{\gamma, S}=Z_{\gamma, S}^{-1}\exp\{-\beta \gamma H_{S}\}\exp\{-\beta\widetilde{\Delta}_{S}\},\label{rSa2}
\end{equation}
\begin{equation}
Z_{\gamma, S}=\textrm{Tr}_{S}(\exp\{-\beta \gamma H_{S}\}\exp\{-\beta\widetilde{\Delta}_{S}\}).\label{rS11}
\end{equation}
Then $Z_{\gamma, S}$ and $F_{\gamma, S}=-\ln (Z_{\gamma, S})/\beta$  become
$\gamma$-dependent and, we obtain the expressions
\begin{equation}
\left\langle H_{S}\right\rangle =\left.-\frac{1}{\beta}\frac{d(\ln Z_{\gamma, S})}{d\gamma}=\frac{dF_{\gamma, S}}{d\gamma}\right|_{\gamma=1}
\end{equation}
\begin{equation}
 C_{S}=\left.\frac{d^{2}F_{\gamma, S}}{d\gamma dT}\right|_{\gamma=1}.\label{alla1}
\end{equation}

\subsection{Approach II (Partition function approach) }\label{aproachII}

The reduced distribution (\ref{rSa}) remains unchanged if we introduce
a certain (possibly temperature-dependent) function $\Upsilon(T)$
and redefine the system partition function and the bath-induced operators
as 
\begin{equation}
Z_{S}\rightarrow Z_{S}/\Upsilon(T),\,\,\,\Delta_{S}\rightarrow\Delta_{S}+\frac{1}{\beta}\ln\Upsilon(T).\label{trans}
\end{equation}
The analogues transformation for the distribution (\ref{rSaF}) reads
\begin{equation}
F_{S}\rightarrow F_{S}-\frac{1}{\beta}\ln\Upsilon(T),\,\,\,\Delta_{S}\rightarrow\Delta_{S}+\frac{1}{\beta}\ln\Upsilon(T).\label{trans1}
\end{equation}
The transformations (\ref{trans}) and (\ref{trans1}) shift the origin
of the bath-induced interaction $\Delta_{S}$. For example, we can
take $\Upsilon(T)=Z_{B}$ (the bath partition function is defined
via Eq. (\ref{Zb})). This choice is especially reasonable for the
weak system-bath coupling, because it makes $\Delta_{S}=0$ if $H_{SB}=0$.
In general, for any $H_{SB}\neq0$, it yields the system partition
function  
\begin{equation}
Z_{S}=Z/Z_{B}.\label{ZZ}\end{equation}
 The definition (\ref{ZZ}) is the key to the approach II. According
to the recipe developed in Refs.\ \onlinecite{for85,for88,for07}, we should
identify (\ref{ZZ}) with the system partition function, and use the
standard Eqs. (\ref{Hd})-(\ref{Cd}) to calculate the necessary thermodynamic
quantities.

The approaches I and II
result in different expressions for thermodynamic quantities,
provided the system-bath coupling is not weak. This will be illustrated
in Sec.\ 4 based on different examples. 
The validity and predictions of the two approaches are discussed in Sec.\ 5.

\section{Analysis of the two different approaches for 
illustrative examples}

\subsection{Model system}

We consider a general system-bath problem.
We assume that the system (e.g., a (macro)molecule) consists of $N_{S}$
point particles, $X_{i}$, $P_{i}$, and $M_{i}$ being their positions,
momenta, and masses. The bath comprises $N_{B}$ point particles
 with positions $x_{i}$, momenta $p_{i}$, and masses $m_{i}$. 
All interactions ($S-S$, $S-B$, and $B-B$) are
pairwise, so that the parts of the overall Hamiltonian 
$H = H_{S} +  H_{B} + H_{SB}$
are explicitly written as follows
\begin{subequations}
\begin{equation}
H_{S}=\sum_{i=1}^{N_{S}}\left\{ \frac{P_{i}^{2}}{2M_{i}}+U_{S}(X_{i})\right\} +\sum_{i>j}^{N_{S}}U_{SS}(X_{i}-X_{j}),\label{HS}
\end{equation}
\begin{equation}
H_{B}=\sum_{i=1}^{N_{B}}\frac{p_{i}^{2}}{2m_{i}}+\sum_{i>j}^{N_{B}}U_{BB}(x_{i}-x_{j}),\label{HB}
\end{equation}
\begin{equation}
H_{SB}=\sum_{i=1}^{N_{S}}\sum_{j=1}^{N_{B}}U_{SB}(X_{i}-x_{j}).\label{HSB}
\end{equation}
\end{subequations}
Here $U_{SS}$, $U_{BB}$, and $U_{SB}$ are the corresponding interaction
potentials (which may be different for any pair of particles $i$
and $j$) and the system is allowed to be subjected to an external
potential $U_{S}$. For clarity, we consider a one-dimensional ensemble.
A generalization to the three-dimensional case is straightforward.

\subsection{Classical mechanics}

We first consider the two approaches to subensemble thermodynamics for
a system of classical point particles.
In classical mechanics, the partition function for any canonical distribution
is a product of the momentum and coordinate contributions. Furthermore,
the momentum contributions to the reduced distribution (\ref{rSa})
can be integrated out, so that the only nontrivial part of the distribution
is  the contribution of the potential energy.

\subsubsection{A single Brownian particle }\label{421}

Let us first consider a single Brownian particle, which corresponds
to $N_{S}=1$ in Eq. (\ref{HS}). After the insertion of Hamiltonians
(\ref{HS}), (\ref{HB}), and (\ref{HSB}) into Eq. (\ref{rS}), we
can make use of the isotropy of space and change the integration variables
$x_{i}\rightarrow x_{i}-X$. This way, we obtain $\Delta_{S}=0$ and arrive
at the standard result that the reduced distribution (\ref{rSa})
is the canonical distribution (\ref{rS1}) determined by the system Hamiltonian
\begin{subequations}
\begin{equation}
\rho_{S}=Z_{S,c}^{-1}\exp\{-\beta(\frac{P^{2}}{2M}+U_{S}(X)\},\label{rSs}
\end{equation}
\begin{equation}
Z_{S,c}=\sqrt{2\pi M/\beta}\int dX\exp\{-\beta U_{S}(X)\}.\label{Zss}
\end{equation}
\end{subequations}
It is tempting to assume that all thermodynamic characteristics of
the Brownian particle can be obtained through the bath-independent,
canonical system distribution (\ref{rSs}) or, what is equivalent,
through the differentiation of the partition function (\ref{Zss})
according to the standard Eqs. (\ref{Hd})-(\ref{Cd}). This is indeed
the case if we use approach I. If we follow approach, however, we
obtain
\begin{equation}
Z/Z_{B}=\eta Z_{S,c}.\label{zz}
\end{equation}
Here $Z_{S,c}$ is the free-particle partition function (\ref{Zss})
and 
\begin{equation}
\eta=\frac{\int dx_{1}...dx_{N_{B}}\,\exp\left\{ -\beta\left(\sum_{i>j}^{N_{B}}U_{BB}(x_{i}-x_{j})+\sum_{j=1}^{N_{B}}U_{SB}(x_{j})\right)\right\} }{\int dx_{1}...dx_{N_{B}}\,\exp\left\{ -\beta\left(\sum_{i>j}^{N_{B}}U_{BB}(x_{i}-x_{j})\right)\right\} }.\label{Heta}
\end{equation}
Apparently, $\eta\neq1$, in general. This becomes evident, e.g.,
if we expand the numerator in Eq. (\ref{Heta}) in powers of $U_{SB}$.
Symbolically, $\eta=1+O(\left\Vert U_{SB}\right\Vert )$. In approach
II, the factor of $\eta$ induces (unphysical) bath-dependence
of the system mean energy, entropy, and the specific heat. 
Thus,  approaches I and II lead, in general, to different predictions
even for a single classical Brownian particle. 

An important exception
is a Brownian particle bilinearly coupled to a harmonic bath, 
\begin{subequations}
\begin{equation}
H_{B}=\sum_{i=1}^{N_{B}}\left(\frac{p_{i}^{2}}{2m_{i}}+\frac{m_{i}\omega_{i}^{2}x_{i}^{2}}{2}\right),\label{HBh}
\end{equation}
\begin{equation}
H_{SB}=\sum_{i=1}^{N_{B}}\frac{m_{i}\omega_{i}^{2}}{2}
\left(X^{2}-2Xx_{i}\right),\label{HSBh}
\end{equation}
\end{subequations}
where $\omega_{i}$ denotes the frequencies of the bath oscillators. In
this case, we obtain $\eta=1$ and the two approaches give the same result. 
This is only the case for the simple form of the bilinear system-bath
coupling. 

If we retain the harmonic bath
(\ref{HBh}) but add a nonlinear interaction term to $H_{SB}$, the situation
differs. Let us consider, for example, the potential 
\begin{equation}
H_{SB}=\sum_{i=1}^{N_{B}}\left\{ \frac{m_{i}\omega_{i}^{2}}{2}\left(X^{2}-2Xx_{i}\right)+\frac{\xi_{i}}{\left(X-x_{i}\right)^{2}}\right\} \label{HSBa}
\end{equation}
where $\xi_{i}$ denote the corresponding constants.
The additional term in the potential, 
the form of which has been chosen for demonstrative purposes, may describe repulsion of the particles at short distances.
 The reduced system
partition function $Z_{S}$ is  given by Eq. (\ref{Zss}).
Incorporating Eqs.\ (\ref{HBh}) and (\ref{HSBa}) into Eq.\ (\ref{Heta}),
we obtain for the factor $\eta$, which describes
the deviation of the system partition function $Z_S$ from the ratio $Z/Z_B$, 
\begin{equation}
\eta=\prod_{j=1}^{N_{B}}\exp\{-\beta\omega_{j}\sqrt{2m_{j}\xi_{j}}\}.
\end{equation}
Thus, even for this rather simple example, the factor $\eta$
can significantly differ from unity and also acquire a temperature dependence.
Within approach II, this would result in incorrect
predictions for $\left\langle S_{S}\right\rangle $, 
$\left\langle H_{S}\right\rangle $,
and $C_{S}$ \cite{foot1}.

\subsubsection{A harmonic dumbbell}\label{422}

Let us suppose that the system (e.g., a molecule) consists of
a collection of point particles. In general, the explicit evaluation
of the bath-induced potential $\Delta_{S}$ beyond the weak system-bath
coupling limit is a difficult task \cite{roux,ben71,gel99}, except in the case
of a harmonic bath (modelling, e.g., a Gaussian solvent) 
bilinearly coupled to the system. In the latter
case, the integrations over $x_{i}$ in Eq.\ (\ref{rS}) can easily
be performed analytically. It is instructive to consider the simplest
nontrivial situation, when the system consists of two identical particles
($N_{S}=2$). Such a model can describe, for example, a diatomic molecule 
or a dumbbell.
If we require the total (harmonic) Hamiltonian $H$ (\ref{H1})
to be translationally invariant, we  arrive at the expression
\begin{subequations}
\begin{equation}
H_{S}=\frac{P_{1}^{2}}{2M}+\frac{P_{2}^{2}}{2M}+\frac{M\Omega_{S}^{2}}{2}(X_{1}-X_{2})^{2},\label{HS2}\end{equation}
\begin{equation}
H_{SB}+H_{B}=\sum_{i=1}^{N_{B}}\left(\frac{p_{i}^{2}}{2m_{i}}+\frac{m_{i}\omega_{1i}^{2}}{2}(X_{1}-x_{i})^{2}+\frac{m_{i}\omega_{2i}^{2}}{2}(X_{2}-x_{i})^{2}\right),\label{HI2}
\end{equation}
\end{subequations}
where $\omega_{1i}$, $\omega_{2i}$, and $\Omega_{S}$ denote the corresponding
oscillator frequencies \cite{foot4}. 

For later use it is convenient to rewrite Eq. (\ref{HI2}) in the 
equivalent form
\begin{equation}
H_{SB}+H_{B}=\sum_{i=1}^{N_{B}}\left(\frac{p_{i}^{2}}{2m_{i}}+\frac{m_{i}}{2}(\omega_{1i}^{2}+\omega_{2i}^{2})\left\{ x_{i}-\frac{\omega_{1i}^{2}X_{1}+\omega_{2i}^{2}X_{2}}{\omega_{1i}^{2}+\omega_{2i}^{2}}\right\} ^{2}\right)+\frac{M\Omega_{\Delta}^{2}}{2}(X_{1}-X_{2})^{2}.\label{HI2W}
\end{equation}
Here 
\begin{equation}
\Omega_{\Delta}^{2}=\sum_{j=1}^{N_{B}}\frac{m_{j}\omega_{1j}^{2}\omega_{2j}^{2}}{M(\omega_{1j}^{2}+\omega_{2j}^{2})}\label{WS2W}
\end{equation}
is the frequency of the bath-induced harmonic potential. 
Integrating out the bath modes,
we obtain a reduced distribution $\rho_{S}$ of the form of 
Eq.\ (\ref{rSa}) with
\begin{subequations}
\begin{equation}
\Delta_{S}=\frac{M\Omega_{\Delta}^{2}}{2}(X_{1}-X_{2})^{2},\label{WS2}
\end{equation}
\begin{equation}
Z_{S}=\frac{2\pi M}{\beta}L\sqrt{\frac{2\pi}{\beta M(\Omega_{\Delta}^{2}+\Omega_{S}^{2})}}.\label{Zdu}
\end{equation}
\end{subequations}
Here $L$ is the (one-dimensional) system volume. It is seen that the
influence of the bath manifests itself in the additional attractive
harmonic potential $\Delta_{S}$, which is coordinate-dependent but
temperature-independent, so that $d\Delta_{S}/dT=0$ \cite{foot4a}.

For the Hamiltonian (\ref{HS2})-(\ref{HI2}), the ratio of the total
and the bath partition functions yields the system partition function,
i.e. $Z/Z_{B}=Z_{S}$ (Eq. (\ref{Zdu})). One might thus
expect that approaches I and II give  the same predictions for the 
thermodynamics quantities. Due to the presence of the bath-induced
potential $\Delta_{S}$, this is, however, not the case. 
Indeed, if we would follow  approach II, 
the mean system energy is given by
\begin{equation}
E_{S}=-\frac{d(\ln Z_{S})}{d\beta}=-\frac{d(\ln \{ Z/Z_{B} \})}{d\beta}=\frac{3}{2\beta},\label{APi}
\end{equation}
 which corresponds to the thermal energy of a system with three degrees
of freedom (one for the center of mass translation and two for the
vibration). If, on the other hand, according to approach I, 
we associate $E_{S}$
with $\left\langle H_{S}\right\rangle $, then Eq.\ (\ref{alla}) yields
\begin{equation}
\left\langle H_{S}\right\rangle =-\frac{d(\ln Z_{S})}{d\beta}-\left\langle \Delta_{S}\right\rangle =\frac{1}{2\beta}\left(3-\frac{\Omega_{\Delta}^{2}}{\Omega_{\Delta}^{2}+\Omega_{S}^{2}}\right).\label{H2corr}
\end{equation}
Thus, approach II predicts for the heat capacity $C_{S}=3k_{B}/2$,
irrespective of the strength of the dumbbell ($\Omega_{S}^{2}$) and
solvent-induced ($\Omega_{\Delta}^{2}$) potentials. This seems to be 
physically incorrect, given that the reduced distribution $\rho_{S}$ 
contains the  bath-induced attractive
harmonic potential $\Delta_{S}$ (\ref{WS2}) and does not coincide 
with the canonical distribution (\ref{rS1})  for the dumbbell alone.
 Approach I, on the other hand, predicts
the bath-dependent specific heat to be 
\begin{equation}
C_{S}=\frac{k_{B}}{2}\left(3-\frac{\Omega_{\Delta}^{2}}{\Omega_{\Delta}^{2}+\Omega_{S}^{2}}\right).\label{Ccorr}
\end{equation}
The coupling to the solvent has the strongest influence 
in case of two free
Brownian particles ($\Omega_{S}^{2}=0$). In this case, the actual number
of degrees of freedom is reduced by one, which is in accordance with 
the physical expectations.

\subsection{Quantum mechanics}

We next analyze the two approaches to subensemble thermodynamics
for quantum mechanical point particles. In the quantum mechanical case,
 there exist
pitfalls and subtleties in the calculation of the specific heat 
already for a single Brownian particle bilinearly coupled to a heat bath of
harmonic oscillators.
This has been demonstrated in Refs.\ \onlinecite{nie02,han06,han08}. 
To elucidate the nature of these subtleties
and to simplify the presentation, we restrict ourselves to a semiclassical
analysis and calculate the leading order ($\sim\hbar^{2}$) 
quantum corrections to the thermodynamic quantities. 
To this end, we employ the Wigner representation \cite{wig84}. 

To simplify the notation, 
we introduce a collective index $a$ which runs over all $N_{S}+N_{B}$
system and bath  particles and use a tilde to denote the
corresponding  positions, momenta, and masses. Thus, the
total Hamiltonian (\ref{H1}) reads
\begin{equation}
H=\sum_{a}\left\{ \frac{\tilde{p}_{a}^{2}}{2\tilde{m}_{a}}\right\} +U(\tilde{x}_{1},...,\tilde{x}_{N_{S}+N_{B}}),\label{W0}
\end{equation}
Within the Wigner representation, we treat $\tilde{p}_{a}$ and
$\tilde{x}_{a}$ as (semi)classical phase space variables. The Hamiltonian
retains its classical form, but the canonical distribution 
for the overall system 
(\ref{ro}) is given by the corresponding Wigner distribution (denoted by the
superscript $W$)
\begin{subequations}
\begin{equation}
\rho^{W}=Z^{-1}\exp\{-\beta H+\lambda H^{(1)}\}+O(\lambda^{2}),\label{rWgen}
\end{equation}
\begin{equation}
Z=\textrm{Tr}(\exp\{-\beta H+\lambda H^{(1)}\})+O(\lambda^{2}),\label{ZWgen}\end{equation}
\begin{equation}
\lambda\equiv(2\pi\hbar)^{2}.\label{lam}
\end{equation}
\end{subequations}
Here, as in the classical case, $\textrm{Tr}(...)$ denotes the integration
over the corresponding phase space variables. The quantum correction
$H^{(1)}$ is explicitly given as \cite{wig84,wig32}
\begin{equation}
H^{(1)}=\sum_{a}\left\{ -\frac{\beta^{2}}{8\tilde{m}_{a}}\frac{\partial^{2}U}{\partial\tilde{x}_{a}^{2}}+\frac{\beta^{3}}{24\tilde{m}_{a}}\left(\frac{\partial U}{\partial\tilde{x}_{a}}\right)^{2}\right\} +\sum_{a,b}\frac{\beta^{3}\tilde{p}_{a}\tilde{p}_{b}}{24\tilde{m}_{a}\tilde{m}_{b}}\frac{\partial^{2}U}{\partial\tilde{x}_{a}\partial\tilde{x}_{b}}.\label{W1}
\end{equation}
It is noted  that $H^{(1)}$  is explicitly temperature-dependent
and contains mixed coordinate-momenta terms.

The reduced Wigner distribution of the system, $\rho^{W}_{S}$, 
is obtained by integrating  Wigner distribution of the overall system, Eq.\ 
(\ref{rWgen}),
over the phase space variables of the bath. 
Since $H$ and $H^{(1)}$ in Eq. (\ref{rWgen}) 
are  functions but not operators,  expressions 
(\ref{rS})-(\ref{2a}) derived in Sec.\ 3
remain also correct for the reduced Wigner distribution. 
In general, the Wigner transform of an operator differs from its 
respective classical expression, because quantum mechanically positions and 
momenta do not commute. 
However, if an operator can be split into a
part which depends only on coordinates and a part which depends 
only on momenta, then
the Wigner transform  is given by the 
corresponding classical expression. 
This is the case for $H_{S}$, $H_{B}$, and $H_{SB}$. 
Therefore, we can use the classical expression for $H_{S}$ while 
evaluating the mean system energy. 


\subsubsection{A quantum Brownian particle}

As in Ref.\ \onlinecite{han08}, 
we consider a single Brownian particle bilinearly
coupled to a heat bath of harmonic oscillators. The system Hamiltonian
is given by Eq.\ (\ref{HS}) with $N_{S}=1$, the bath Hamiltonian
by Eq.\ (\ref{HBh}) and the system-bath coupling by Eq. (\ref{HSBh}).
Inserting the corresponding formulas into Eqs.\ (\ref{W0}) and (\ref{W1})
we can  integrate the bath degrees of freedom out of the overall
(system+bath) Wigner distribution (\ref{rWgen}) and arrive at the
system distribution 
\begin{equation}
\rho_{S}^{W}=Z_{S}^{-1}\exp\{-\beta(H_{S}+\Delta_{S})\}+O(\lambda^{2}).\label{rSW}
\end{equation}
Here $H_{S}=P^{2}/(2M)$ is the free particle Hamiltonian, the
bath-induced interaction operator reads
\begin{equation}
\Delta_{S}=-\lambda\frac{P^{2}}{2M}\frac{(\Omega\beta)^{2}}{12},\,\,\,\Omega^{2}=\sum_{j=1}^{N_{B}}\frac{m_{j}\omega_{j}^{2}}{M},\label{HSW}
\end{equation}
and the partition function is given by the expression
\begin{equation}
Z_{S}=L\sqrt{\frac{2\pi M}{\beta(1-\lambda(\Omega\beta)^{2}/12)}}=L\sqrt{\frac{2\pi M}{\beta}}\left(1+\lambda(\Omega\beta)^{2}/24)\right)+O(\lambda^{2}).\label{zW}
\end{equation}
The bath-induced operator $\Delta_{S}$ is  position-independent,
but depends on momentum and temperature. 

The ratio of the total ($Z)$ and bath ($Z_{B}$) partition functions
can also be readily obtained from expressions (\ref{rWgen})-(\ref{W1})
to yield
\begin{equation}
Z/Z_{B}=L\sqrt{\frac{2\pi M}{\beta}}\left(1-\lambda(\Omega\beta)^{2}/24)\right)+O(\lambda^{2}).\label{zWR}
\end{equation}
A comparison of Eqs.\ (\ref{zWR}) and (\ref{zW}) shows that the partition function of the system is not given by the ratio of the total ($Z)$ and bath ($Z_{B}$) partition functions. The quantum
corrections in the two expressions
have the same magnitude but opposite signs. 

Since the bath-induced operator $\Delta_{S}$ is momentum and temperature
dependent, it is not expected that the differentiation of the partition
function alone gives the averaged energy $\left\langle H_{S}\right\rangle $.
Indeed, the calculation gives
\begin{equation}
-d(\ln Z_{S})/d\beta=\frac{1}{2\beta}-\frac{\lambda\Omega^{2}}{12}\beta+O(\lambda^{2}).\label{zWa}
\end{equation}
On the other hand, if we follow the approach I and use Eq. (\ref{alla}) 
or perform
directly an average over the distribution (\ref{rSW}), we obtain the 
correct value
\begin{equation}
\left\langle H_{S}\right\rangle =\frac{1}{2\beta}+\frac{\lambda\Omega^{2}}{24}\beta+O(\lambda^{2}).\label{HsW}
\end{equation}
Finally, approach II predicts for the average system energy 
\begin{equation}
E_{S}=-d(\ln\{ Z/Z_{B}\})/d\beta=\frac{1}{2\beta}+\frac{\lambda\Omega^{2}}{12}\beta+O(\lambda^{2}).\label{zWRa}
\end{equation}
It is noted that Eqs.\ (\ref{zWRa}) and (\ref{HsW}) have been 
derived in \cite{han08} using a different method. 
The results obtained for the average system energy 
via the different approaches, Eqs.\ (\ref{zWa}), (\ref{HsW}), and (\ref{zWRa}),
are all different, thus providing a nice example of how noncritical use of
the standard thermodynamic equations (\ref{alla}) can lead to ambiguous
results.

\subsubsection{A quantum harmonic dumbbell}

As a final example we consider the harmonic model 
of a dumbbell coupled to a bath quantum mechanically \cite{dumb1,dumb2}.
 The thermodynamics of the quantum 
harmonic dumbbell  can  be constructed
within the Wigner distribution method employing the general equations
(\ref{rWgen})-(\ref{W1}). The total classical dumbbell+bath Hamiltonian $H$
in Eq. (\ref{rWgen}) is given by Eqs. (\ref{HS2}) and (\ref{HI2W}).
Further, $H$ is inserted 
into Eq. (\ref{W1}) to derive the quantum correction $H^{(1)}$. The so-obtained expression is, however,
quite cumbersome and is not presented. Here, we only give the results necessary for the discussion of the thermodynamics quantities.

The reduced Wigner distribution of the system is given by the general 
formula Eq.\ (\ref{rSW}), where $H_{S}$ is defined via Eq. (\ref{HS2}) and
the bath induced operator reads
\begin{equation}
\Delta_{S}=\Delta_{S}^{0}-\frac{\lambda\beta^{2}}{12}\Delta_{S}^{1}.\label{HSWd}\end{equation}
Here we have introduced the notation
\begin{subequations}
\begin{equation}
\Delta_{S}^{0}=\frac{M\Omega_{\Delta}^{2}}{2}(X_{1}-X_{2})^{2},\label{HSWd0}\end{equation}
 \begin{equation}
\Delta_{S}^{1}=M(\Omega_{S}^{2}+\Omega_{\Delta}^{2})^{2}(X_{1}-X_{2})^{2}+\frac{P_{1}^{2}}{2M}(\Omega_{1}^{2}+\Omega_{S}^{2})+\frac{P_{2}^{2}}{2M}(\Omega_{2}^{2}+\Omega_{S}^{2})-\frac{P_{1}P_{2}}{M}\Omega_{S}^{2},\label{HSWd1}
\end{equation}
\begin{equation}
\Omega_{1}^{2}=\frac{1}{M}\sum_{j=1}^{N_{B}}m_{j}\omega_{1j}^{2},\,\,\,\Omega_{2}^{2}=\frac{1}{M}\sum_{j=1}^{N_{B}}m_{j}\omega_{2j}^{2}.\label{WS2d}
\end{equation}
\end{subequations}
and the frequency  $\Omega_{\Delta}^{2}$ of the bath-induced harmonic potential is defined via Eq. (\ref{WS2W}).
The bath-induced operator $\Delta_{S}$ is explicitly coordinate-dependent
(due to the classical contribution (\ref{HSWd0})) as well as momentum
and temperature-dependent (due to the quantum correction (\ref{HSWd1})).
The corresponding partition function reads 
\begin{equation}
Z_{S}=Z_{S}^{cl}\left\{ 1+\frac{\lambda\beta^{2}}{24}\left(4\Omega_{S}^{2}+2\Omega_{\Delta}^{2}+\Omega_{1}^{2}+\Omega_{2}^{2}\right)\right\},\label{Zdud}
\end{equation}
where the classical system partition function $Z_{S}^{cl}$ is given 
by Eq. (\ref{Zdu}). 

The mean energy of the dumbbell  calculated via approach I 
(employing Eq.\ (\ref{rSW}))
is given by the expression
\begin{equation}
\left\langle H_{S}\right\rangle =\left\langle H_{S}\right\rangle _{cl}+\lambda\left\langle H_{S}\right\rangle _{q}+O(\lambda^{2}).\label{HsWd}
\end{equation}
Here the first term is the classical contribution,
\begin{equation}
\left\langle H_{S}\right\rangle _{cl}=\frac{1}{\beta}\left(1+\frac{1}{2}\frac{\Omega_{S}^{2}}{\Omega_{\Delta}^{2}+\Omega_{S}^{2}}\right)\label{HsWdC}
\end{equation}
and the quantum correction reads 
\begin{equation}
\left\langle H_{S}\right\rangle _{q}=\frac{\beta}{24}\left\{ 5\Omega_{S}^{2}+\Omega_{1}^{2}+\Omega_{2}^{2}+\frac{1}{2}\frac{\Omega_{S}^{2}}{\Omega_{\Delta}^{2}+\Omega_{S}^{2}}\left(2\Omega_{S}^{2}+\Omega_{1}^{2}+\Omega_{2}^{2}\right)\right\} .\label{HsWdQ}
\end{equation}
 In the limit
 $\Omega_{S}^{2}=0$, $\Omega_{1}^{2}=\Omega_{2}^{2}$ the dumbbell 
reduces to two noninteracting Brownian particles. Correspondingly, the mean
energy Eq.\ (\ref{HsWd}) gives twice of what is predicted by Eq.\ (\ref{HsW}). 
The same is true for the bath induced interaction operators $\Delta_S$ (cf.\ 
Eqs.\ (\ref{HSWd}) and (\ref{HSW})).

On the other hand, if we follow approach II, we obtain for the ratio of the partition functions of the overall system and the bath
 \begin{equation}
Z/Z_{B}=Z_{S}^{cl}\left\{ 1-\frac{\lambda\beta^{2}}{24}\left(2\Omega_{S}^{2}+\Omega_{1}^{2}+\Omega_{2}^{2}\right)\right\} \label{zWRd}
\end{equation}
and correspondingly as prediction for the mean energy of the system
 \begin{equation}
-d(\ln\{ Z/Z_{B}\})/d\beta=\frac{3}{2\beta}+\frac{\lambda\beta}{12}\left(2\Omega_{S}^{2}+\Omega_{1}^{2}+\Omega_{2}^{2}\right)+O(\lambda^{2}).\label{zWRad}
\end{equation}
The comparison with Eq.\ (\ref{HsWd}) shows that  approaches I
and II give not only different classical contributions to the mean system
energy, but also very different quantum corrections.
It is also interesting to note that the strength of the
bath-induced potential, $\Omega_{\Delta}^{2}$, does not enter 
the expression for the mean energy,
Eq.\ (\ref{zWRad}), while the
mean energy calculated via approach I (Eq.\ (\ref{HsWd})) depends
sensitively on this quantity.

\section{Discussion and Conclusions}

The results presented above demonstrate that the two different approaches
to describe the thermodynamics of a subsystem can predict very different 
results if the system-bath coupling is not weak. This was already shown
earlier for a quantum harmonic oscillator \cite{han06} and for a quantum
Brownian particle \cite{han08} bilinearly coupled to a harmonic bath.
The results obtained here corroborate and extend these earlier findings. 
The study also shows that ambiguities in the description of reduced 
thermodynamics already occur in the classical case for more complex 
systems, such as an anharmonic bath (Sec.\ \ref{421}) or if
 the system under study consists of more than a single point particle 
(Sec.\ \ref{422}). 
Based on the results above we shall now analyze the two approaches and discuss
their advantages and shortcomings.

The different predictions of the two approaches for the
thermodynamic quantities
can be related to the ambiguities in the definition of
the energy of a system that is coupled to the environment.
If the system-bath coupling is not negligible, the system-bath interaction
$H_{SB}$ (or a certain part of it) may be included in  the system 
energy \cite{ced98} thus resulting in a variety of definitions.
 Furthermore, the
results for the thermodynamics quantities may depend on the particular
physical quantity that is considered fundamental (e.g. the system energy or
the partition function).

Approach I associates the mean system energy $E_{S}$ with the
expectation value of the system Hamiltonian $\left\langle H_{S}\right\rangle $.
The definition $E_{S}=\left\langle H_{S}\right\rangle $
clearly associates the observable ($E_{S}$) with the corresponding
physical operator $H_{S}$. This definition appears natural and fits
into the general scheme of the statistical thermodynamics and probability
theory. For example, if we think of $E_{S}$ as the mean internal 
energy (e.g.,
as of the internal energy of a molecule with several vibrational degrees
of freedom), the choice $E_{S}=\left\langle H_{S}\right\rangle $
is well physically justified. As a result of the definition 
$E_{S}=\left\langle H_{S}\right\rangle $, the
bath and the system-bath coupling influence
the mean system energy $E_{S}$ only indirectly, 
through the reduced distribution (density matrix) of the system $\rho_{S}$
given by Eq.\ (\ref{rSa}). It is important to note that $\rho_{S}$ 
does not coincide with the  canonical
distribution for the isolated system.  As a consequence, the
thermodynamics of the reduced system is described by the relations
(\ref{alla})-(\ref{2a}). Once the definition 
$E_{S}=\left\langle H_{S}\right\rangle $
is accepted, no other assumptions are necessary to construct the system
thermodynamics. Corroborating the results 
obtained in Refs.\ \onlinecite{han06,han08}, the present  extended
study shows that the use of  this definition gives rise
to physically and logically consistent results for both quantum mechanical and
classical systems. 

The fundamental quantity of approach II, on the other hand, is
the partition function of the system, $Z_{S}$. 
To obtain thermodynamics quantities, approach II involves two
steps: First,  the partition function of the system is identified as
the ratio of the total and bath partition functions $Z/Z_{B}$ 
(Eq.\ (\ref{rSa})). In addition to
this choice, approach II assumes that the standard thermodynamical relations
given by Eqs.\ (\ref{Hd})-(\ref{Cd}) can be used to calculate 
thermodynamics quantities such as the mean energy, the entropy, and the specific heat. 
 The choice $Z_S = Z/Z_{B}$ for the partition function of the system 
appears to be  reasonable,
 notably in the limit of weak system-bath coupling. However, as discussed 
in Sec.\ \ref{aproachII}, the partition function $Z_{S}=Z/Z_{B}$ corresponds the reduced
distribution $\rho_{S}$ given in Eqs.\ (\ref{rSa}), (\ref{trans}), 
which does not coincide with
the corresponding canonical distribution for the isolated system alone. 
Instead, $\rho_{S}$
contains an additional bath-induced operator $\Delta_{S}$ which,
in general, depends on the temperature and on the degrees of
freedom of the system. In such a case, as has been shown in 
Sec.\ \ref{genexpr},  the thermodynamic relations for a subensemble, 
Eqs.\ (\ref{alla})-(\ref{2a}), should be employed instead of the
standard thermodynamic relations (\ref{Hd})-(\ref{2}), provided the system-bath
coupling is not small. 
Therefore, there is no any \textit{a priori}
theoretical justification for plugging $Z_{S}=Z/Z_{B}$ into Eqs.
(\ref{Hd})-(\ref{2}), and we have
to additionally postulate that differentiations of $F_{S}=-\ln(Z/Z_{B})/\beta$
give us, according to Eqs. (\ref{Hd})-(\ref{Cd}), the mean system
energy, entropy, and the specific heat.
 Thus the two fundamental assumptions of approach II, 
the choice for $Z_S$ and the validity of
the standard thermodynamic relations (\ref{Hd})-(\ref{2}) even for the 
subensemble, can not be proven within the approach
itself.


It also worthwhile to mention another peculiarity of approach II.
As has been pointed out in Refs.\ \onlinecite{han06,han08}, the mean energy 
of the system obtained within approach II corresponds to the definition
\begin{equation}
E_{S}=\left\langle H\right\rangle -\left\langle H_{B}\right\rangle _{B}\equiv\left\langle H_{S}\right\rangle +\left\langle H_{B}+H_{SB}\right\rangle -\left\langle H_{B}\right\rangle _{B},\label{ESii}
\end{equation}
where $\left\langle ...\right\rangle _{B}$ denotes averaging over the bath
distribution (Eq.\ (\ref{Zb})). 
For nonvanishing system-bath interaction, the term 
$\left\langle H_{B}+H_{SB}\right\rangle -\left\langle H_{B}\right\rangle _{B}$
gives an additional contribution to $E_S$ that is not present in 
approach I. As a consequence of the structure of this additional term,
it is not possible  to introduce
an operator of the mean energy, whose average will give  $E_{S}$. 
Furthermore, considering for example $E_{S}$ as the mean internal
energy of a molecule with several vibrational degrees of freedom,
it does not appear to be consistent that the definition (\ref{ESii}) contains
contributions which are explicitly determined by the bath degrees
of freedom.

To summarize, although   approaches I and II give  identical result
for the thermodynamics of a subsystem if the system-bath coupling 
is negligible, 
their  predictions differ significantly for finite system bath coupling.
These differences arise because  different quantities are considered
as fundamental in the two approaches and be related to
the different definitions of the mean energy of the system used.  
The results obtained above and those presented earlier \cite{han08} suggest
that approach I is superior both from the physical and 
the logical point of view.

\begin{acknowledgments}
This work has been supported by the Deutsche Forschungsgemeinschaft (DFG) through 
the DFG Cluster of Excellence ``Munich Centre of Advanced Photonics'' 
(M. F. G.) and by the Fonds der Chemischen Industrie (M. T.).
M. F. G. is deeply grateful to Daniel Kosov, Rainer H\"artle, and Wolfgang Domcke 
 for useful discussions.  
\end{acknowledgments}


\begin{thebibliography}{10}
\bibitem{foot3}To avoid notational abuse, hereafter we drop the subscript $S$ in
denoting the system averages because, even strictly speaking, 
$\left\langle Y_{S}\right\rangle _{S}\equiv\left\langle Y_{S}\right\rangle$
for any $Y_{S}$ which depends on the system degrees of freedom only.
\bibitem{foot1}It is noted that approach II has been constructed and 
tested only for a quantum harmonic oscillator
bilinearly coupled to a harmonic bath \cite{for85,for88,for07} and, to our cknowledge, has never been applied to more complex potentials.
\bibitem{foot4}In the spirit of normal mode analyses \cite{str1,str2} we have taken
$\omega_{1i}\neq\omega_{2i}$ even for symmetric dumbbell, because
its two constituting particles may experience different local potentials
from the bath molecules.
\bibitem{foot4a}$\exp\left\{ -\beta\Delta_{S}\right\} $ is a close analogue of the
pair correlation function for the bath molecules.
\bibitem{dumb1}It was only very recently 
when the exact master equation for the quantum dumbbell was derived and solved \cite{dumb2}.
\end{thebibliography}
\end{document}